\documentclass[12pt]{article}
\usepackage{graphicx}

\newcommand{\blackslug}{\mbox{\hskip 1pt \vrule width 4pt height 8pt 
depth 1.5pt \hskip 1pt}}
\newcommand{\QED}{\quad\blackslug\lower 8.5pt\null\par\noindent}

\title{Linear Programming helps solving large multi-unit combinatorial auctions}
\author{Rica Gonen\thanks{Supported by Grant 15561-1-99 from the Israel
Ministry of Science, Culture and Sport} \and Daniel Lehmann 
\\ School of Computer Science and Engineering, \\
Hebrew University, Jerusalem 91904, Israel }

\begin{document}
\date{April 20th 2001}
\maketitle
\begin{abstract}
Previous works suggested the use of Branch and Bound techniques for finding the optimal
allocation in (multi-unit) combinatorial auctions. They remarked that Linear Programming
could provide a good upper-bound to the optimal allocation, but they went on using lighter
and less tight upper-bound heuristics, on the ground that LP was too time-consuming 
to be used repetitively to solve large combinatorial auctions.
We present the results of extensive experiments solving large (multi-unit) combinatorial auctions
generated according to distributions proposed by different researchers.
Our surprising conclusion is that Linear Programming is worth using. 
Investing almost all of one's computing time in using LP to bound from above the value
of the optimal solution in order to prune aggressively pays off.
We present a way to save on the number of calls to the LP routine 
and experimental results comparing different heuristics for choosing the bid to be considered next.
Those results show that the ordering based on the square root of the size of the bids that
was shown to be theoretically optimal in a previous paper by the authors performs surprisingly
better than others in practice. Choosing to deal first with the bid with largest coefficient
(typically 1) in the optimal solution of the relaxed LP problem, is also a good choice.
The gap between the lower bound provided by greedy heuristics and the upper bound provided
by LP is typically small and pruning is therefore extensive.
For most distributions, auctions of a few hundred goods among a few thousand bids can
be solved in practice. All experiments were run on a PC under Matlab.
\end{abstract}

\section{Background, Introduction and Previous Work}
\label{sec:intro}
The last decade has seen extensive denationalizations and a fundamental change in economic
patterns brought about by the use of the Internet as a world-wide market place.
Both phenomena promoted the use of auctions to a central place among economic mechanisms.
The study of complex auctions is now a most active area of research, drawing from 
the fields of Mechanism Design in Economics and from the Theory of Algorithms in Computer Science.
Combinatorial auctions in which a large number of items are sold and in which 
bidders may express preferences over bundles, i.e., subsets, of items have drawn a lot of attention
recently~\cite{RothPekHar, Sandholm99, Sandholm, Sandholm:DSS, Sandholm:AAAI99, FLBS:99,
LCS:EC99, Nisan:EC00, GonLeh:EC00, deVriesVohra:survey, Camus2:AAAI, Tenn:sometractauc}.

We consider here the ``winner determination'' problem, i.e., the algorithmic problem of finding 
the allocation that maximizes the declared social welfare.
We consider both (single-unit) combinatorial auctions and multi-unit combinatorial auctions
in which a number of identical items of each type are sold.
In~\cite{RothPekHar}, this problem was shown to be NP-hard.
In~\cite{Sandholm99, LCS:EC99}, it is shown that it is even hard to find a non-trivial approximation
to the optimal allocation. Those results consider worst-case complexity and do not necessarily
imply that, in practice, one cannot find the optimal solution of real-life large combinatorial
auctions. Some of the papers above proposed winner-determination algorithms and tried to
show they perform well even on large auctions. Others proposed mechanisms based on non-optimal
allocations.
A number of researchers proposed sets of problems on which to test candidate algorithms:
most notably Leyton-Brown, Shoham and Tennenholtz~\cite{Camus2:AAAI}, Leyton-Brown, 
Pearson and Shoham~\cite{TestSuite:EC00} and de Vries and Vohra~\cite{deVriesVohra:survey}.

We implemented a Branch-and-Bound search program along the lines of~\cite{GonLeh:EC00} for 
finding the optimal allocation in multi-unit combinatorial auctions. This program makes
heavy use of a linear programming routine to bound from above the value of extensions
of partial allocations.
Common wisdom in the field has it that linear programming is too time-consuming to be used in practice,
and one should use lighter, if if less accurate, methods to obtain upper-bounds.
We report here results that go against this common wisdom: pruning is of the essence and one
should not hesitate to devote all the time necessary to compute tight upper-bounds
in order to prune aggressively.
Such conclusions hold for all the distributions described above.
We present, in Section~\ref{subsec:save}, an important way of saving on those expensive calls to
the LP routine.
Section~\ref{sec:BB} presents the algorithm we used and the remainder of the paper discusses
our experimental results.

\section{Branch and Bound}
\label{sec:BB}
\subsection{Winner Determination in Multi-Unit Combinatorial Auctions}
\label{subsec:WD}
We assume $n$ different commodities and $k_{j}$ identical items of commodity
$j$ for \mbox{$j=1, \ldots n$}.
The problem we are trying to solve is that of determining the optimal
allocation of the items.
A set of bids is given: each bid requests a number of units (possibly zero) 
from each commodity and offers a price $p$ for the whole set.
A subset of the set of all bids is conflict-free if, for each commodity,
the sum of the units requested does not surpass the number of units for sale.
The problem is to find a conflict-free subset that maximizes
the sum of the prices proposed. 
\subsection{Our Algorithm}
\label{subsec:alg}
The algorithm we experimented with is very similar to the one described in~\cite{GonLeh:EC00}.
We assume that none of the bids submitted requests more units than available, i.e.,
each singleton set of bids is conflict-free.
Our algorithm consists of an initialization phase and a main (recursive) routine.

In the initialization phase, we use a host of fast heuristics to find a {\em good} allocation.
In practice, this initialization phase is performed very fast and gives an allocation that
is not much less than the optimal one. Many times, one, in fact, gets the optimal allocation,
but one, obviously, does not know it is optimal.
The heuristics used are all members of the greedy family: one chooses a bid $b$, grants it,
subtracts the quantities requested by $b$ from the stock of available units, eliminates all
bids that cannot be granted anymore, i.e., those that are not conflict-free, chooses a bid,
and so on. Each heuristic provides a feasible allocation. One keeps the best one.
The different heuristics we propose to use differ only in the way they choose the next bid to be granted.
The best allocation found is kept.

The main (recursive) routine is given:
\begin{enumerate}
\item a partial allocation, i.e., a set of bids already granted,
\item a stock of units available, 
\item a set of bids not yet granted, each of them requesting only a number of units not larger than
the number of units in stock, and
\item it uses a global variable describing the best allocation found so far.
\end{enumerate}
The problem is solved by calling the main routine with an empty partial allocation,
the full stock of available units, the set of all bids, 
and after having initialized the best allocation to the
best allocation found in the initialization phase.

The main routine is:
\begin{enumerate}
\item \label{empty} {\bf Stop} if the third argument is empty, i.e., no more bids waiting, return,
\item \label{update} {\bf Update} if the value of the partial allocation is larger 
than that of the best solution found so far,
update this best solution,
\item \label{bound} {\bf Bound} bound from above the value of the optimal allocation of the units 
left in stock to the bids that are the third argument of the routine. 
This is done by solving the Linear Programming
problem that is the fractional relaxation of the Integer Programming problem describing the
optimal allocation,
\item \label{prune} {\bf Prune} checks whether the search can be stopped, i.e., 
the subtree rooted at the current node can 
be pruned. The condition for pruning is that the sum of the values of the partial allocation
and the solution of the Linear Programming problem just found be not greater than the value
of the best allocation found so far. If the condition holds, return. If the condition does not
hold, go on to the next step,
\item \label{choose} {\bf Choose} choose a bid from the third argument,
\item \label{left} {\bf Left Call} call recursively the main routine with a partial allocation
augmented by the bid just chosen, the stock of available units diminished by subtracting the units
requested by the bid chosen and the list of bids diminished by subtracting the chosen bid
and all bids that request more units than available in (the diminished) stock,
\item \label{right} {\bf Right Call} call recursively the main routine with the original
partial allocation, the original stock and the list of bids without the chosen bid.
\end{enumerate}

The {\bf Left Call} solves the auction obtained after the chosen bid has been granted.
The {\bf Right Call} solves the auction obtained after the chosen bid has been denied.

The only step under-specified in this routine is {\bf Choose}: we shall discuss at length
how the next bid to branch on should be chosen in Section~\ref{sec:order}.
The only time-consuming step in the main routine is the Linear Programming step in
{\bf Bound}. We have a way to save on those expensive calls to the LP routine.

\subsection{Saving on calls to LP}
\label{subsec:save}
\begin{figure}[htbp]
\centering
\includegraphics[height=8cm]{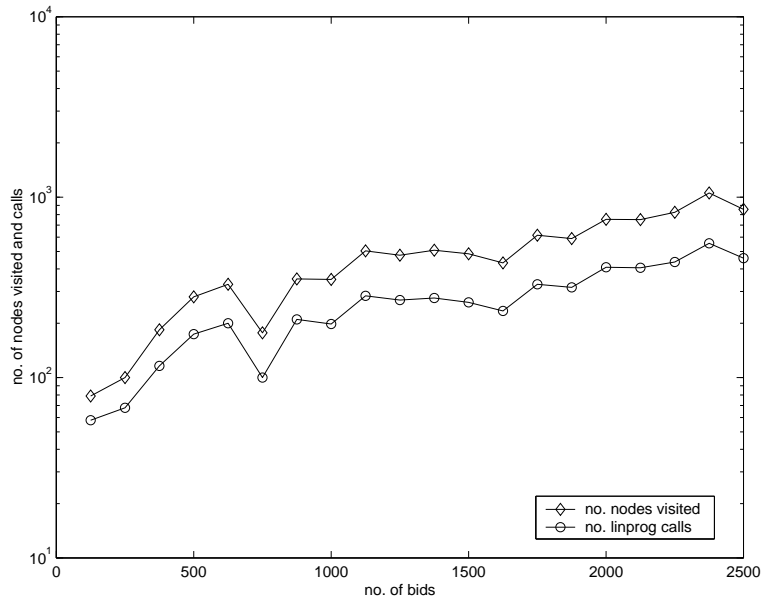}
\caption{Number of nodes visited vs. Number of calls to the LP routine}
\label{fig:nodes_linprogcalls}
\end{figure}

Notice that, in {\bf Right Call}, the problem to be solved is very similar to the original one:
same partial allocation, same stock. This recursive call will trigger a call to the LP routine
(call it LP2) solving a LP problem that is very similar to the one just solved (call it LP1). 
The only difference is
that the bid just chosen has disappeared from the list of bids considered by LP2.
If this bid did not enter the optimal (fractional) solution of LP1, i.e., if its coefficient
in the solution of LP1 was equal to zero, then, the solution of LP2 is the same as that of LP1
and therefore one can save a call to the LP routine. One can easily see that, in this case,
no pruning can take place.
Figure~\ref{fig:nodes_linprogcalls} shows that the saving can be considerable.
The auctions are taken from~\cite{Camus2:AAAI}. 
On the x-axis: the number of bids considered,
on the y-axis: under log scale, the number of nodes visited or the number of calls to the LP routine.

\section{Experimental Parameters}
\label{sec:expparam}
We ran all our experiments on a Pentium Pro 930 MHz with 512K of memory, running Windows 2000.
Our programs were coded in Matlab (R12). We used the linprog facility of the Optimization ToolBox
for Linear Programming: this is a primal-dual algorithm.
All our results are averages over fifteen auctions drawn from a distribution to be
specified for each graph.
We did experience some memory problems (probably due to some remaining bugs in Matlab Release 12),
but found Matlab extremely easy to use and its linprog facility excellent.
The remainder of the paper presents, in graphic form, the results of our experiments.
Some of those graphs are difficult to decipher in black and white: we apologize and
will produce more readable graphs for the conference paper.
\section{Experimental Results}
\label{sec:results}
\subsection{Global Picture}
\label{subsec:global}
\begin{figure}[htbp]
\centering
\includegraphics[height=8cm]{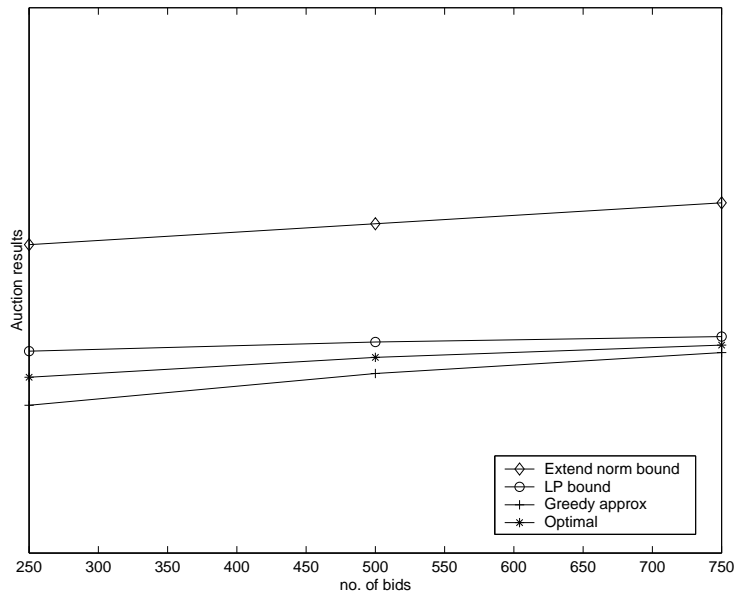}
\caption{Optimal solution vs. Lower Bound and two Upper Bounds}
\label{fig:gr_opt_lp}
\end{figure}

\begin{figure}[htbp]
\centering
\includegraphics[height=8cm]{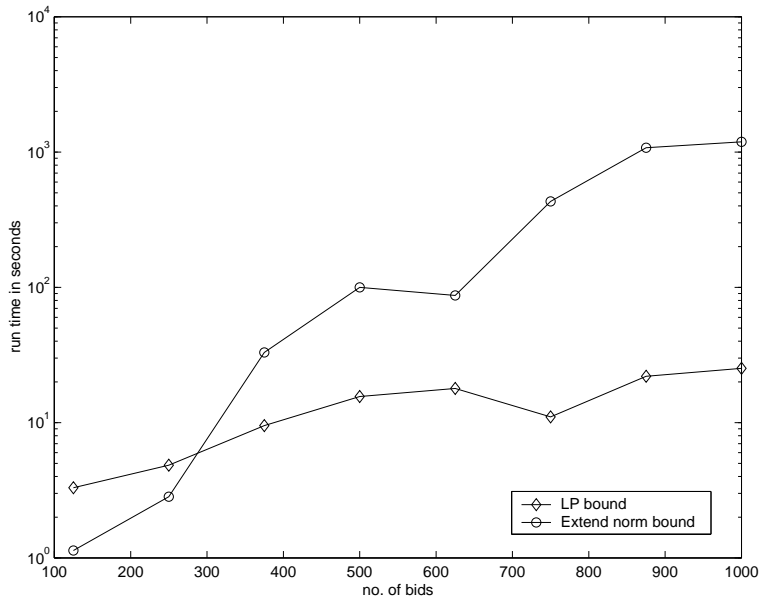}
\caption{Running Times using LP vs. Leyton-Brown's upper bound}
\label{fig:bounds}
\end{figure}

Figure~\ref{fig:gr_opt_lp} describes the quality of the optimal solution, 
the lower bound obtained in the initialization phase
and two upper bounds for multi-unit auctions.
The upper curve (extended norm bound) describes the easy to compute upper bound 
proposed in~\cite{Camus2:AAAI}. 
The second (from the top) curve describes the upper bound given by Linear Programming.
The latter can be proved to be always at most equal to the former.
On the x-axis: the number of bids considered (from 250 to 750) 
and the y-axis the average value (in monetary terms,
numbers are not important) of the allocations.
This graph shows two things. First, that the lower bound computed in the initialization phase and
the upper bound computed by LP are very close. The optimal solution is, obviously, in-between.
The larger the auction, the better the convergence of those three values.
The second thing is that the easy-to-compute upper-bound of~\cite{Camus2:AAAI} is not even close
to the LP bound and seems to get (relatively) worse as the number of bids increases.
The main claim of this paper is that using the time-consuming LP routine is preferable to the
use of a lighter but easier to compute upper bound in {\bf Bound}.
Figure~\ref{fig:bounds} compares the running times of the algorithm above that uses LP and
the running time of the same algorithm using the easy-to-compute upper bound proposed by Leyton-Brown
and al.
The distribution of auctions is taken from~\cite{Camus2:AAAI}, for auctions of different sizes.
On the x-axis: the number of bids. On the y-axis: the running time in seconds (log scale).
By using LP, one gains two orders of magnitude for large auctions.  
The lighter upper bound is better for very small numbers of bidders.

\section{Initialization Phase}
\label{sec:iniphase}
\begin{figure}[htbp]
\centering
\includegraphics[height=8cm]{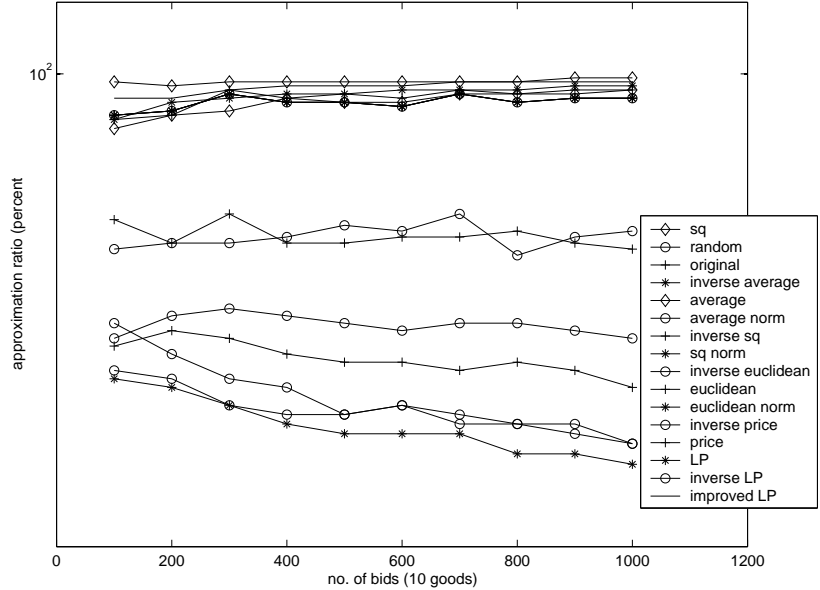}
\caption{Value of different greedy heuristics}
\label{fig:appratio}
\end{figure}

In the initialization phase, in order to find a good solution to be used for pruning,
we propose to use many greedy heuristics and keep the best solution found.
Figure~\ref{fig:appratio} describes the quality of 16 different greedy heuristics on auctions
of varying size (100 to 1000 bids, always 10 goods) taken from the distribution of~\cite{Camus2:AAAI}.
On the y-axis: the percentage of the value of the optimal solution reached by the approximation:
100 means perfect approximation, 50 means half of the optimal solution.
One expects that criteria that pick up more promising bids first give better results.
This is what happens.
One sees three main groups.
The middle group includes two criteria:  picking at random and picking following the
(in fact random) order with which bids were given. 
The lower (worst) group of criteria comprises criteria that choose unattractive bids first.
The higher group contains all criteria we thought attractive. It contains both unnormalized and
normalized criteria (see~\cite{GonLeh:EC00}).
The square-root criterion: 
\begin{equation}
\label{eq:unnorm0}
r(a) = \frac{p}{\sqrt{\sum_{j=1}^{n} q_{j}}} \: ,
\end{equation}
that has been proved to be theoretically optimal, gives the best approximation.
The average price per unit gives markedly inferior results.
Two criteria based on Linear Programming give pretty good results.
The first one solves the relaxed LP problem for the original auction and orders the bids
according to their coefficient in the optimal solution, in descending order.
The second, improved one, is an adaptative version of the first: it solves a LP problem before
choosing each bid (after having eliminated all bids that cannot be granted in full).
More on those criteria is found in Section~\ref{sec:order}.

\section{Choosing the bid on which to branch}
\label{sec:order}
\begin{figure}[htbp]
\centering
\includegraphics[height=8cm]{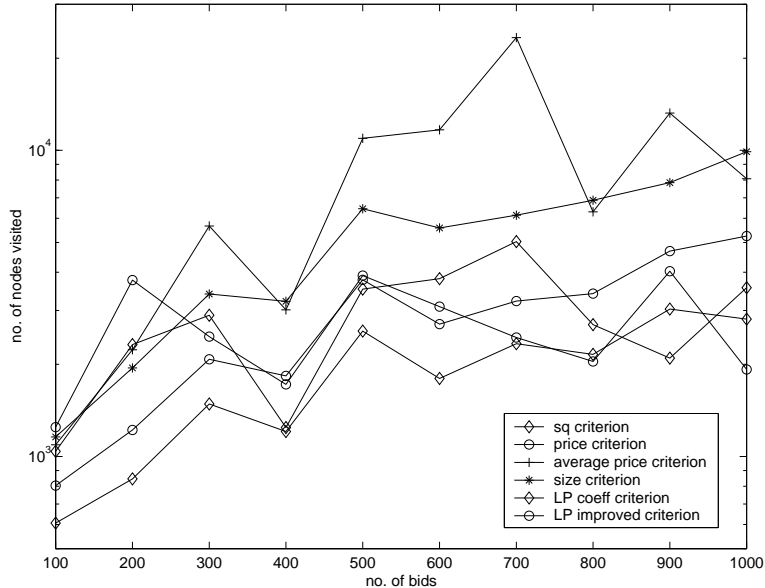}
\caption{Number of nodes visited for different choosing criteria}
\label{fig:sort_100_1000}
\end{figure}

\begin{figure}[htbp]
\centering
\includegraphics[height=8cm]{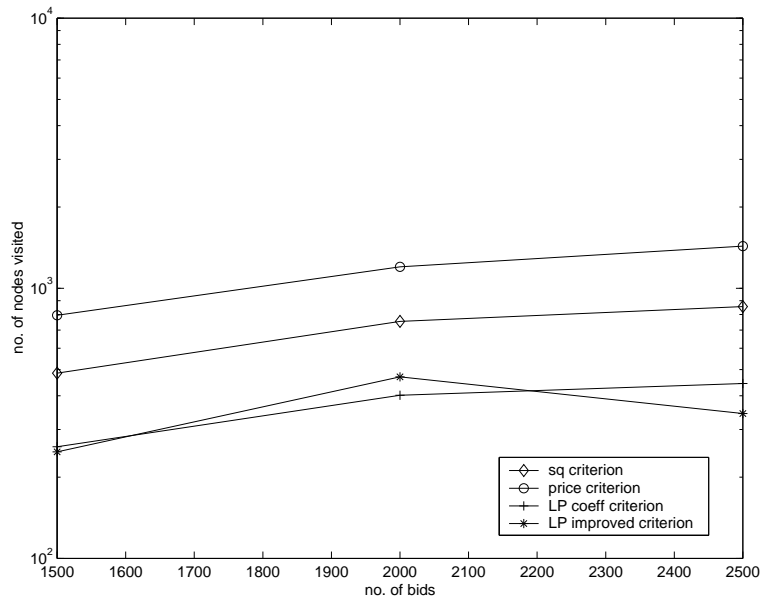}
\caption{Number of nodes visited for different choosing criteria}
\label{fig:sort_1500_2500}
\end{figure}

The main factor in determining the efficiency of our Branch-and-Bound algorithm is the
criterion used in choosing the next bid on which to branch in step {\bf Choose}.
Figure~\ref{fig:sort_100_1000} compares different such criteria.
The graph compares six different criteria.
The auctions were taken from~\cite{Camus2:AAAI}.
The x-axis shows the number of bids and the y-axis, with a log scale, 
shows the average number of the nodes visited.  
The square root criterion of Equation~\ref{eq:unnorm0} seems best. 
Nevertheless the LP coefficient criterion and the LP improved criterion described in 
Section~\ref{sec:iniphase} did very well and even seem better than the square root criterion 
for large numbers of bids.
Figure~\ref{fig:sort_1500_2500} extends the previous graph to larger auctions, for the best criteria
only (for the other criteria we could not finish the search). 
The criteria based on LP are definitely best.

\section{Run Time}
\label{sec:runtime}
\begin{figure}[htbp]
\centering
\includegraphics[height=8cm]{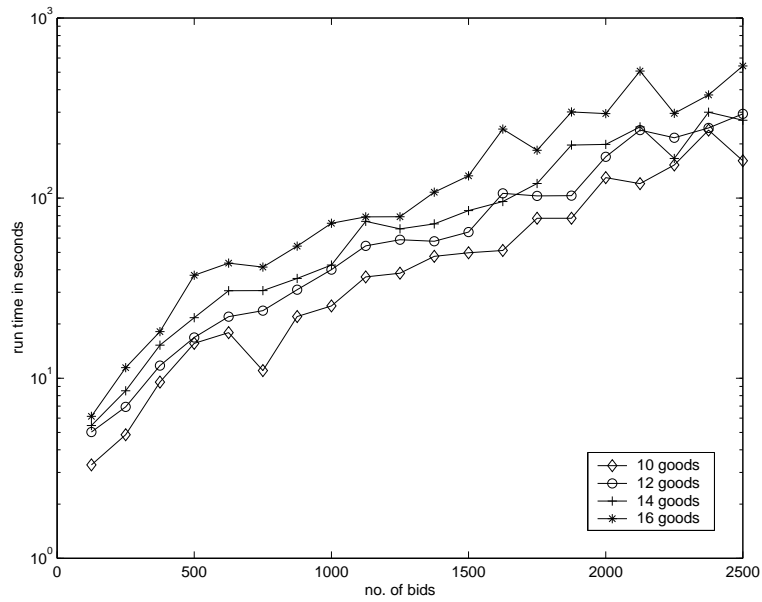}
\caption{Running Times for Leyton-Brown and al.'s distribution}
\label{fig:camus}
\end{figure}

\begin{figure}[htbp]
\centering
\includegraphics[height=8cm]{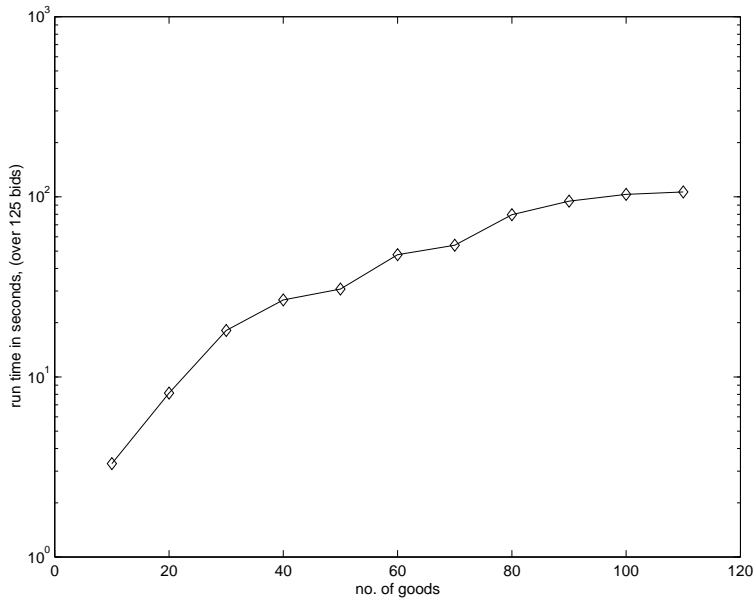}
\caption{Running Times for Leyton-Brown and al.'s distribution}
\label{fig:goods}
\end{figure}

We now describe running times for auctions of different sizes, under different distributions.
Figure~\ref{fig:camus} uses the distribution proposed in~\cite{Camus2:AAAI}.
Auctions have been generated for 10, 12, 14 and 16 goods.
On the x-axis: the number of bids.
On the y-axis: running times in seconds (log scale).
The curves are sub-linear on a logarithmic scale and exhibit the sub-exponential running time of
our algorithm.  
This graph improves substantially on the similar graph found in~\cite{GonLeh:EC00}, showing, again,
the advantage of using LP. 
This graph also shows faster running times for large auctions than the ones 
presented in~\cite{Camus2:AAAI}.
Figure~\ref{fig:goods} presents similar results for a fixed number of bids (125), for different
numbers of goods.
The curve is very clearly sub-linear. 
Contrary to CAMUS, our BB algorithm exhibits a sub-exponential
dependence on the number of goods.

\begin{figure}[htbp]
\centering
\includegraphics[height=8cm]{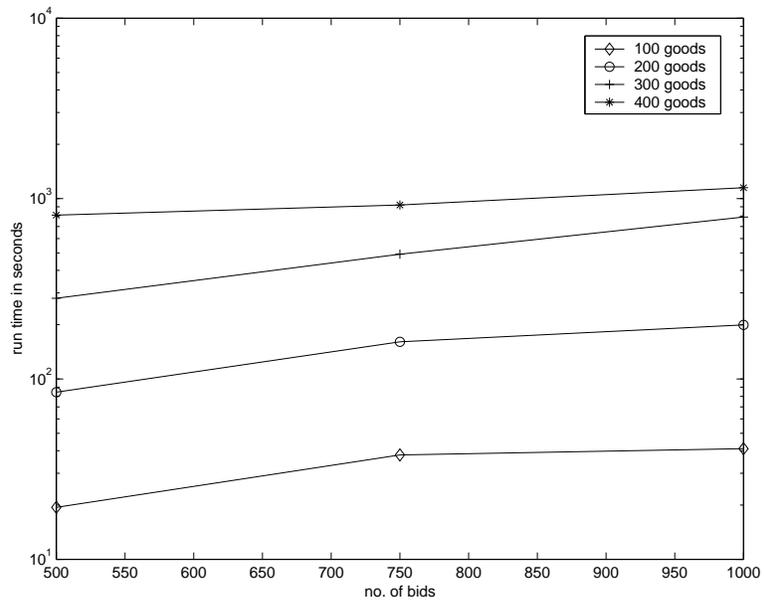}
\caption{Running Times for Sandholm Random distribution}
\label{fig:devries1}
\end{figure}

\begin{figure}[htbp]
\centering
\includegraphics[height=8cm]{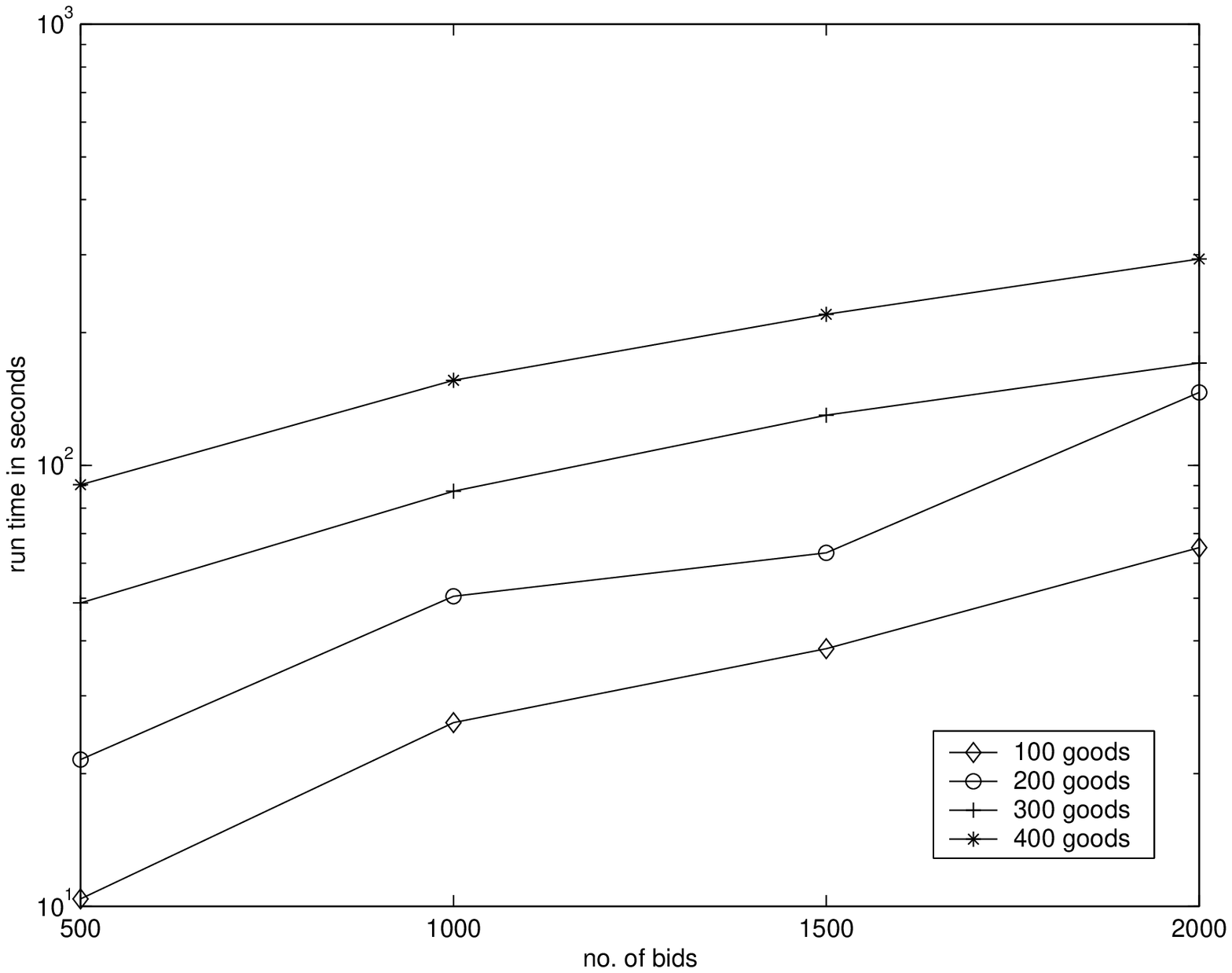}
\caption{Running Times for Sandholm Weighted Random distribution}
\label{fig:devries2}
\end{figure}

\begin{figure}[htbp]
\centering
\includegraphics[height=8cm]{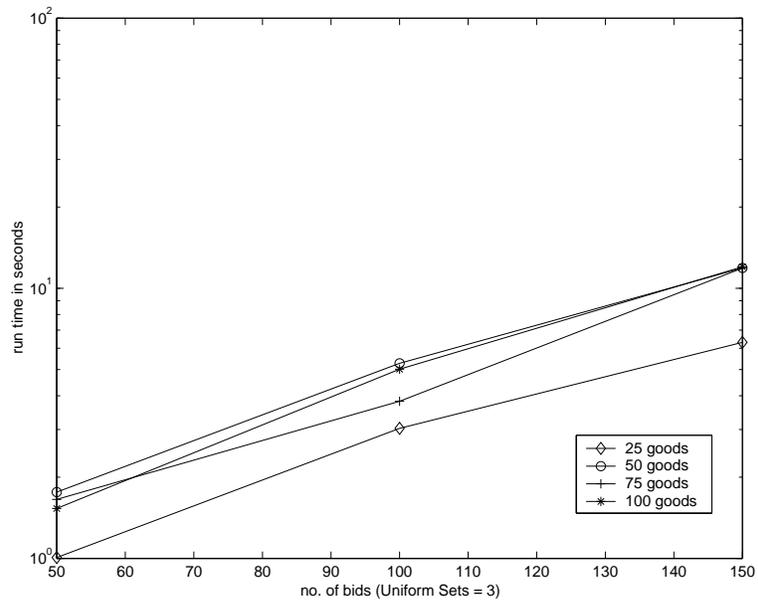}
\caption{Running Times for Sandholm Uniform distribution: sets of size 3}
\label{fig:devries3}
\end{figure}

\begin{figure}[htbp]
\centering
\includegraphics[height=8cm]{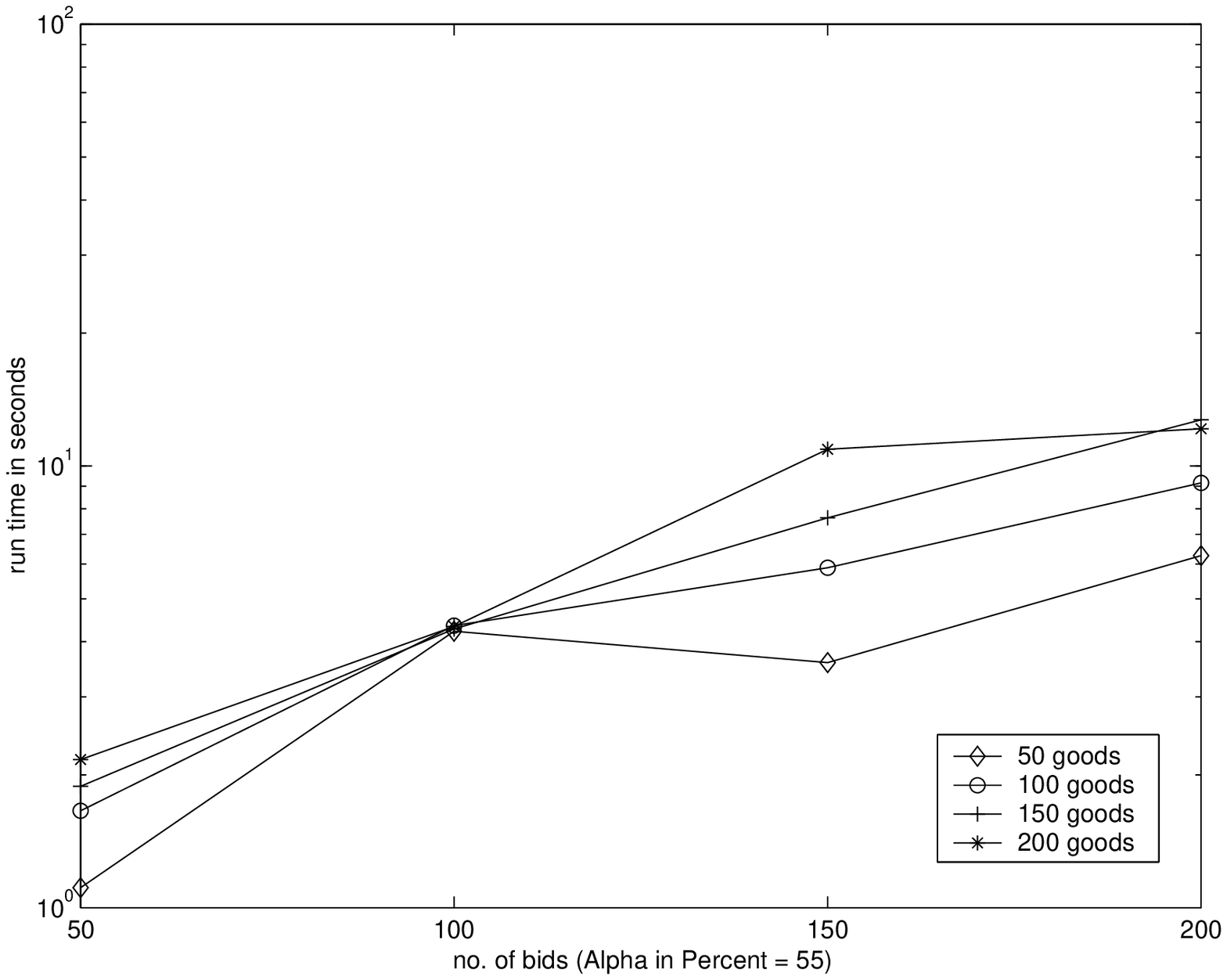}
\caption{Running Times for Sandholm Decay distribution: $\alpha = 55\%$}
\label{fig:devries4}
\end{figure}

Figures~\ref{fig:devries1}, \ref{fig:devries2}, \ref{fig:devries3} and~\ref{fig:devries4} 
describe the running times 
of our algorithm over four distributions described by de Vries and Vohra~\cite{deVriesVohra:survey}.
The auctions generated there are single-unit combinatorial auctions.

\begin{figure}[htbp]
\centering
\includegraphics[height=8cm]{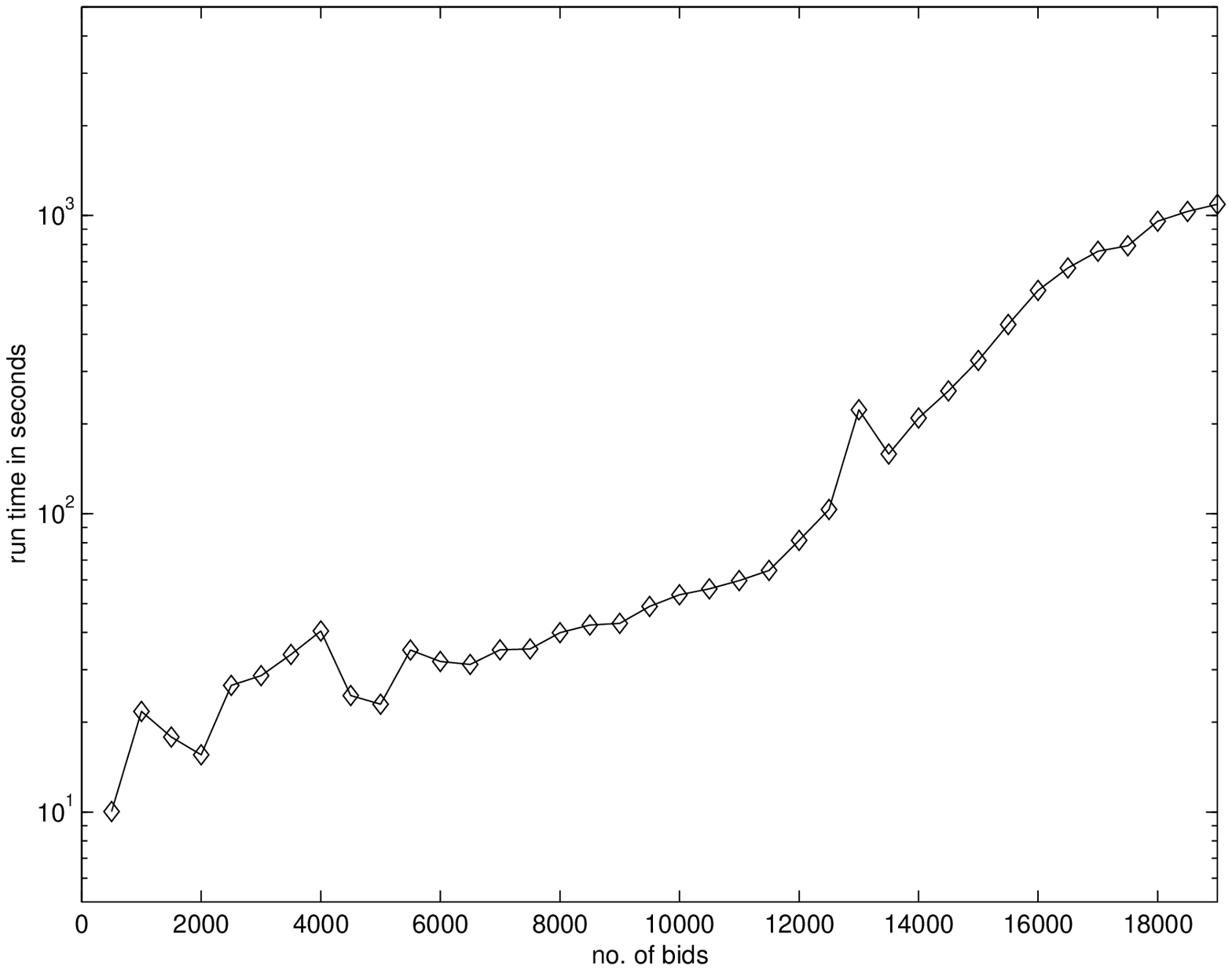}
\caption{Running Times for multipaths distribution}
\label{fig:paths_max}
\end{figure}

\begin{figure}[htbp]
\centering
\includegraphics[height=8cm]{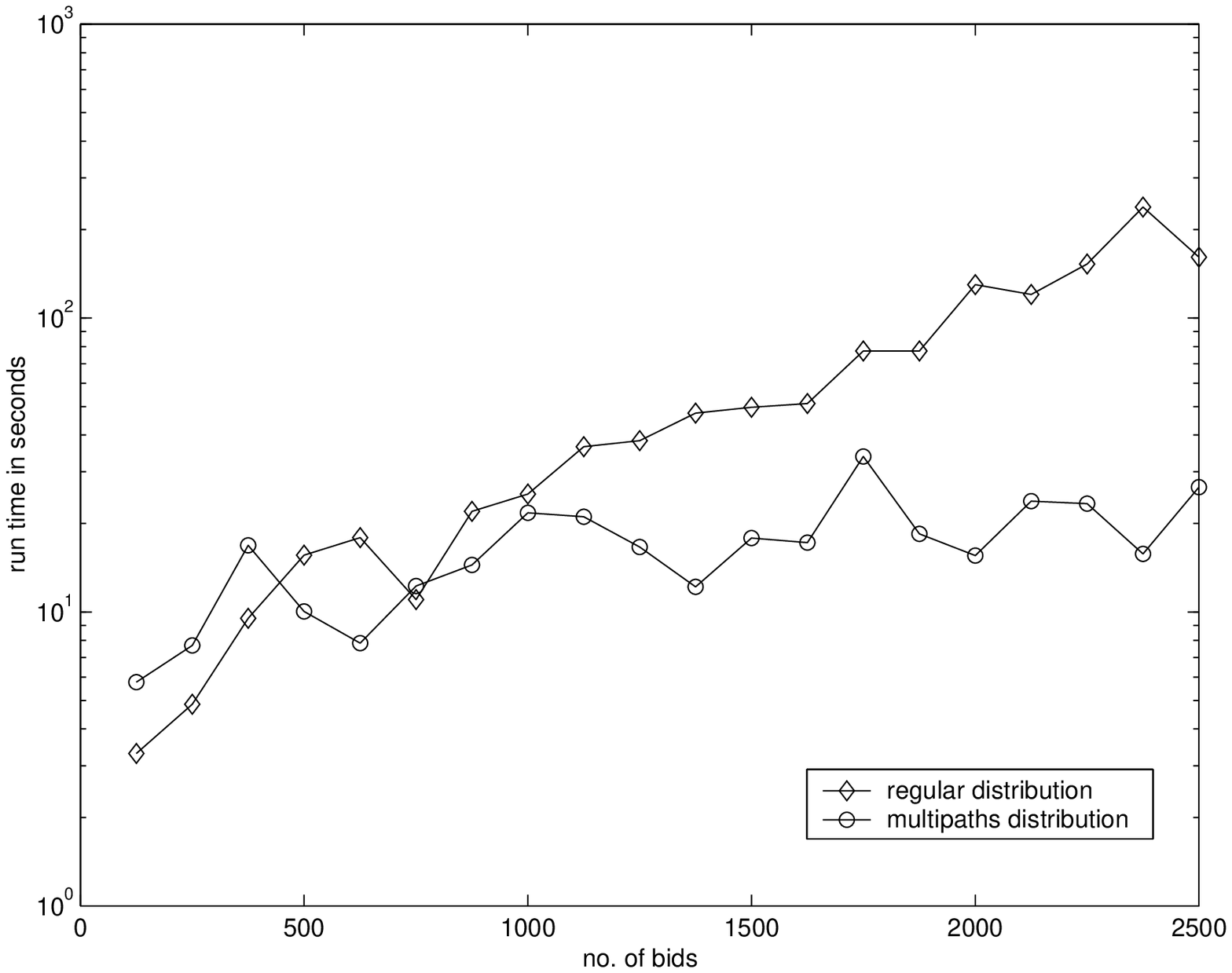}
\caption{Relative Difficulty of Leyton-Brown and al.'s distributions}
\label{fig:pathscomp}
\end{figure}

\begin{figure}[htbp]
\centering
\includegraphics[height=8cm]{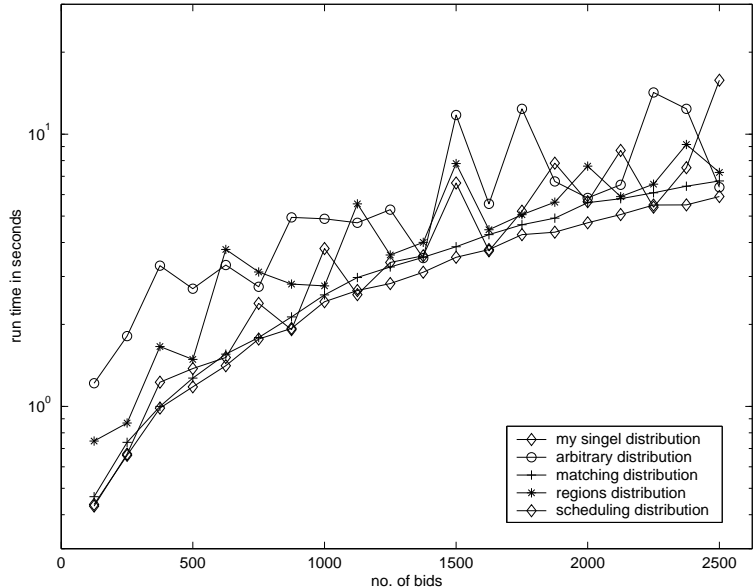}
\caption{Running Time for single-unit CA's}
\label{fig:catssingle}
\end{figure}

Figure~\ref{fig:paths_max} concerns auctions
drawn according to the CATS multipaths distribution of the Test Suite in~\cite{TestSuite:EC00}.
Notice that our algorithm solves quite easily even auctions with up to 20000 bids.
It seems, then, that the CATS multipaths distribution generates problems that are much easier,
on average, than those generated by the distribution previously proposed in~\cite{Camus2:AAAI}.
This claim is supported by~\ref{fig:pathscomp}.
For large numbers of bids, the CATS multipaths distribution indeed generates problems
that are easier to solve.
There is a simple explanation as to why our BB algorithm solves easily CATS auctions
with very large numbers of bids: in fact, one call to the LP routine is typically enough.
What happens is that the initialization phase finds an allocation of value $v$ and the first call
to the LP routine finds a fractional solution of same value $v$.
This means that the LP fractional problem has an integer solution and the initialization
phase found an optimal solution. Our BB algorithm prunes the first node.
Figure~\ref{fig:catssingle} shows the running times of our algorithm for the distributions 
of single unit combinatorial auctions suggested by CATS.  

\section{Conclusions}
\label{sec:concl}
Our first conclusion is that, even though winner determination is theoretically intractable,
it can be performed in practice for auctions of a few hundreds goods among a few thousands
bids, without a super-computer and without low-level programming optimization.
The suggestion of~\cite{LCS:EC99} to consider approximately-efficient
mechanisms, may have been premature.
The challenge is now to tackle auctions of a few thousands goods (such as the FCC auction).
But for such situations, the conceptual framework we have worked with in this paper is 
most probably not adequate:
bidders cannot be expected to describe their preferences by an explicit list of bids.
Suitable languages for expressing preferences must be found and algorithmic problems studied
in such a new framework.

Our second conclusion is that the difficulty in solving combinatorial auctions stems more from
the number of items for sale than from the number of bids. If the number of bids is large,
with respect to the number of items, Linear Programming often finds the optimal integer
solution in one call.

Our third conclusion is that the method of choosing the next bid on which to branch
is extremely important in determining the efficiency of the search.
More work is needed to understand what makes a good choice and why the square-root criterion
of Equation~\ref{eq:unnorm0} and the LP adaptative criterion described in Section~\ref{sec:order} are
so good.
 
\bibliographystyle{abbrv}

\end{document}